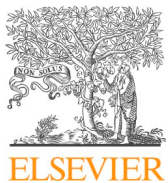
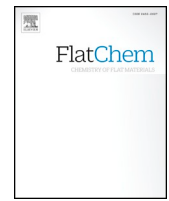

# Single-step exfoliation and functionalization of few-layers black phosphorus and its application for polymer composites

Antonio Esaú Del Río Castillo[a], Christian Daniel Reyes-Vazquez[a,b], Laura Edith Rojas-Martinez[a,b], Sanjay B. Thorat[c], Michele Serri[a], Ana Laura Martinez-Hernandez[b], Carlos Velasco-Santos[b], Vittorio Pellegrini[a,c], Francesco Bonaccorso[a,c,*]

[a] *Graphene Labs, Istituto Italiano di Tecnologia. Via Morego 30, Genoa 16163, Italy*
[b] *Tecnológico Nacional de México/Instituto Tecnológico de Querétaro, División de Posgrado e Investigación, Av. Tecnológico, Querétaro 76150, Mexico*
[c] *BeDimensional SpA, Via Albisola 121, Genoa 16163, Italy*

### ARTICLE INFO

### ABSTRACT

We demonstrate the large scale exfoliation of bulk black phosphorus into few-layers black phosphorus. We use wet-jet mill (WJM) as an industrially scalable exfoliation technique. In order to avoid the degradation by oxidation of the exfoliated flakes, the few-layers black phosphorus is protected by functionalisation (with 4-bromobenzene-diazonium tetrafluoroborate) during the exfoliation process. We used the exfoliated and functionalised flakes as fillers in a polymethyl methacrylate matrix to test the mechanical endurance, obtaining an improvement in the Young's modulus of 106% compared to the bare polymer. These data support the development of black phosphorus materials as 2D-polymer fillers and performance-enhancers.

## 1. Introduction.

Phosphorene, a single layer of phosphorus atoms arranged in a buckled-structure, has remarkable (opto)electronic properties [1–4]. The bulk counterpart of phosphorene, black phosphorus, has a thickness-dependent direct band-gap that can be tuned from 0.3 eV up to 1.8 eV, the latter for single layer, *i.e.*, phosphorene [5]. Bulk blackphosphorus has a carrier mobility of $1000\,cm^2\,V^{-1}\,s^{-1}$ at room temperature [6], making phosphorene and few-layers black phosphorus attractive materials for sensors [7,8], electronic [2,9] and optical devices [10–12]. Furthermore, the structural anisotropy of phosphorene provides two different Young's modulus values according to the direction in which the stress is applied, *i.e.*, along with the zigzag or armchair edges, contrary to the isotropic 1 TPa Young modulus for graphene [13]. The instability of phosphorene under ambient conditions [14] makes the experimental measurement of its Young modulus difficult, but calculations have provided Young's modulus values of 160 and 40 GPa when the stress is applied in the zigzag and armchair directions, respectively [2,15]. Interestingly, the stability of black phosphorus against oxidation increases with the flake thickness, being few-layers black phosphorus and bulk black phosphorus more stable than the single-layer counterpart [16–18]. In this manner, it is thus possible to experimentally measure the Young's modulus of few-layers black phosphorus, obtaining values of 58.6 and 27.2 GPa along the zigzag and armchair axis, respectively [15]. These values make few-layers black phosphorus and phosphorene potential materials to be used as fillers for polymer composites [10,19,20].

The improvement of the mechanical performances of polymer composites with layered materials is one of the most studied and promising fields for the application of such nanomaterials [13,21,22]. To mention an example regarding the mechanical improvement of polymers using layered materials, the Young modulus of polyether ether ketone (PEEK, a techno-polymer known for its high thermal and chemical stability [23], and biocompatibility [24,25], Young's modulus of 3.6 GPa [26] and tensile strength of < 100 MPa [26]) increases of 44% with the addition of graphene nanoplatelets (10% wt) [27]. Such mechanical improvement has opened new perspectives for PEEK applications in bone regeneration and implants [27]. The few-layers black phosphorus and phosphorene used as filler reinforcement are expected to observe similar mechanical improvements in the polymer composites. Actually, the mechanical enhancement at the lab scale was recently demonstrated in polyvinylchloride and poly (vinyl alcohol) and

Conflict of Interest: The authors declare that they have no known competing financial interests or personal relationships that could have appeared to influence the work reported in this paper.

* Corresponding authors at: Graphene Labs, Istituto Italiano di Tecnologia. Via Morego 30, Genoa 16163, Italy (F. Bonaccorso).
*E-mail addresses:* antonio.delrio@iit.it (A.E. Del Río Castillo), francesco.bonaccorso@iit.it (F. Bonaccorso).

https://doi.org/10.1016/j.flatc.2019.100131
Received 14 May 2019; Received in revised form 5 August 2019; Accepted 12 August 2019
Available online 22 August 2019

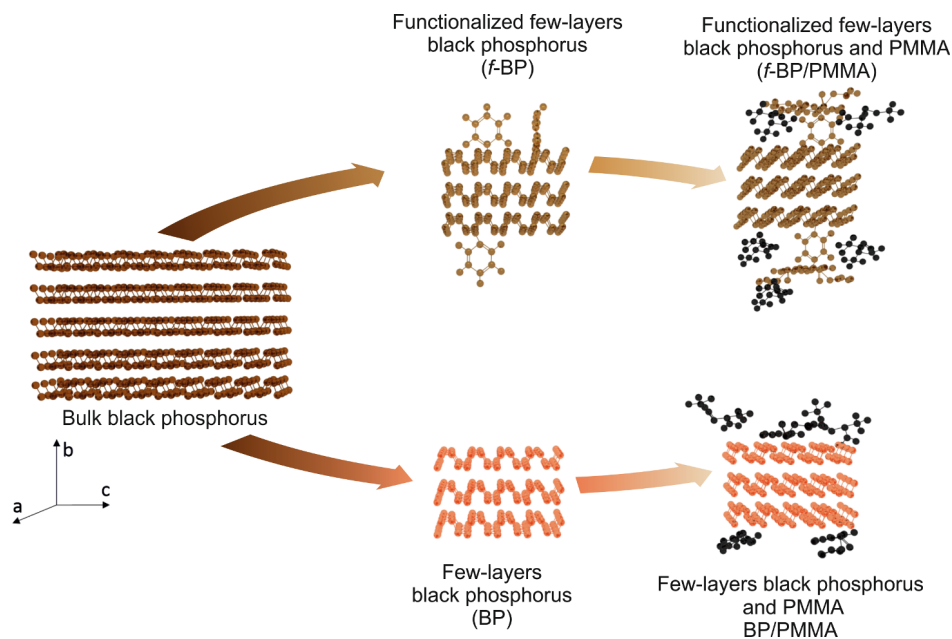

**Fig. 1.** Schematic representation of the bulk black phosphorus, the exfoliated flakes with and without functionalization, and mixed with polymethyl methacrylate.

few-layers black phosphorus composite, almost doubling the Young's modulus of the bare polymer in both cases [28,29].

However, despite the promising proof-of-principle applications of few-layers black phosphorus, the instability under ambient conditions hinders its real exploitations [10,12,30–33]. To make the liquid phase-exfoliated few-layers black phosphorus and phosphorene flakes stable at ambient conditions (i.e., to water and oxygen), it has been proposed to functionalize them [34–36], trap them into a polymer matrix [10,37], or to use high viscous solvents acting as a protecting shell layer on the flakes [29,38]. The stabilisation against oxidation via covalent functionalization, e.g., using diazonium salts, forms C-P bonds between the few-layers black phosphorus and the organic molecules, preventing further oxidation of the exfoliated flakes [34–36]. In fact, the functionalized few-layers black phosphorus flakes have shown no evidence of oxidation after six weeks of exposure in ambient conditions [35]. Following the covalent modification, Cao et al., used the diazonium functionalization to synthesise poly[(1,4-diethynylbenzene)-alt-9,9-bis(4-diphenylaminophenyl) fluorene] grafted onto the few-layers black phosphorus surface [39]. Aryl iodide salts also enable covalent modification and prevent the degradation of few-layers black phosphorus under ambient conditions [34]. Additionally, non-covalent functionalization of black phosphorus has been tested, using 7,7,8,8-tetracyano-p-quinodimethane as a functional group, demonstrating stability against oxidation under ambient conditions [36]. The stabilisation of few-layers black phosphorus in a polymeric matrix has also been demonstrated, for example, by using polycarbonate [10] and polystyrene [37]. In both cases, the trapping of the flakes inside the polymer matrix prevents the degradation of few-layers black phosphorus (as demonstrated by Raman and FTIR spectroscopies) [10,37]. The incorporation of few-layers black phosphorus in a polymer composite is expected to have a two-fold effect: on the one hand, the few-layers black phosphorus could improve the polymer mechanical properties [28,29], while on the other hand, the polymer matrix itself stabilises the few-layers black phosphorus against oxidation [10,37].

The technology transition from the laboratory to the industry has always been a challenge. For materials science, it is of paramount importance to demonstrate the scalability of the synthesis process or the feasibility to exploit the current techniques used in industry for the synthesis of new materials. In this regard, the large scale exfoliation of black phosphorus is carried out in a high-pressure homogeniser: the wet-jet mill (WJM). The WJM process is currently exploited in the ceramic [40], pharmaceutical [41] and food [42] industry to pulverise and homogenise the processed materials [40,41]. The use of WJM for the exfoliation of layered materials, e.g., graphite, boron nitride and transition metal dichalcogenides, has been demonstrated in previous works [43]. The WJM allows the production of high-quality 2D materials dispersions (in the form of single-/few-layer flakes) in litre-scale volumes, at high concentrations (gram per litre-scale) and in short times (second/minute-scale) [44,45]. In this regard, the exfoliation of black phosphorus can be carried out in degassed aqueous [12] or organic solvents, e.g., N-Methyl-2-pyrrolidone – NMP – [5], N-Cyclohexyl-Pyrrolidone – CHP – [10,29], or acetone [46]. For applications in which the presence of solvent residuals is detrimental for the performance, e.g., in electronic devices or polymer composites, the use of easy-to-remove solvents is advantageous [10]. Acetone has a boiling point of 56 °C [47] and is generally considered as a safe solvent [48] (especially if it is compared to NMP or CHP), making it attractive for the aforementioned applications.

In this work, we propose the liquid phase exfoliation of black phosphorus at large scale. Specifically, the WJM process enables continuous and massive production of the exfoliated layered materials [43,44]. To prevent the degradation of the exfoliated flakes during or after the WJM process, the black phosphorus is exfoliated and functionalized with 4-bromobenzene-diazonium tetrafluoroborate in a single step. Finally, we use the exfoliated and functionalized flakes as fillers for polymer composites, demonstrating the mechanical improvement of a polymer-black phosphorus composite. The as-prepared composites show an increase in the Young modulus and tensile strength of 106% and 134%, respectively (compared to the bare polymer), obtained with a concentration of 0.01 wt% of functionalized few-layers black phosphorus. The infrared and Raman spectroscopies show that the composite materials do not show degradation by oxidation for more than eight weeks. Fig. 1 shows a schematic representation of the composites based on few layers- (as control) and functionalized-few layers black phosphorus and polymethyl methacrylate (PMMA).

## 2. Materials and methods

### 2.1. Exfoliation of black phosphorus

250 mg of bulk black phosphorus crystals, from Smart Elements, were ground using a mortar in an inert atmosphere (Ar), to obtain a powder-like material. The powder-like black phosphorus was mixed with 500 mL of anhydrous acetone (Merck), and placed into a 500 mL glass bottle. The exfoliation was performed in a wet-jet mill (Jokoh JN100), using the 0.30 mm nozzle aperture. The piston-pass (defined as the number of times the piston was charged and discharged with solvent/layered material) was set to 50 passes (10 mL per pass). The processed sample was collected in a second container (*i.e.*, a glass bottle). The wet-jet milling process was repeated, passing the acetone-black phosphorus through the 0.20 and 0.10 nozzles consecutivelly. This sample was named BP.

### 2.2. Exfoliation and functionalization of black phosphorus.

250 mg of bulk black phosphorus crystals, from Smart Elements, were ground using a mortar under an inert atmosphere (Ar) obtaining a powder-like material. The powder-like black phosphorus was mixed with 10 mM solution of 4-bromobenzene-diazonium tetrafluoroborate – BBD – (Merck) in 500 mL of anhydrous acetone (Merck) and placed into a 500 mL glass bottle. The exfoliation and functionalization were performed in a wet-jet mill (Jokoh JN100), using the same process followed for the black phosphorus (BP). In this case, the sample was named *f*-BP.

### 2.3. Composite preparations

The PMMA composites were prepared by stirring the BP or *f*-BP dispersions into the PMMA powder, followed by the evaporation of the acetone at room temperature, under a constant nitrogen flow. PMMA powder (2.5 g) was poured into polytetrafluoroethylene Petri dishes, and 5 mL of BP or *f*-BP were added to obtain the 0.01, 0.05, 0.1, 0.2 and 0.5 wt% of exfoliated black phosphorus with respect to PMMA. Besides, samples of bare PMMA were prepared and used as the control. Then the solvent was dried under nitrogen flow at room temperature, followed by drying in a vacuum oven (Binder, VD-53) for two hours. Finally, the obtained films were hot-pressed at two metric tons, at 170 °C for 5 min.

### 2.4. Scanning electron microscopy (SEM).

Pieces of ca. 0.25 cm$^2$ were selected from the films and coated with carbon (thickness: 10 nm). The scanning electron microscopy was performed using an FE-SEM Joel GSM-7500FA, operated at an acceleration voltage of 10 kV, and equipped with an INCA 450 energy dispersive X-ray spectrometer (EDS) and a four-sector backscattered electron detector (BSD). The surfaces were coated with carbon to avoid charging effects. The carbon coating was deposited by using an Emitech sputter coater K550 unit, a K 250 carbon coating attachment, and a carbon cord at a pressure of $1 \times 10^{-2}$ mbar.

### 2.5. Transmission electron microscopy (TEM)

The exfoliated and functionalized black phosphorus flakes were analysed by TEM. Experimentally, 10 μL were diluted in 1 mL of acetone and drop-cast onto an ultrathin formvar polymer film on holey carbon 400 mesh Cu grids (Ted Pella Inc.). The grids were stored under vacuum at room temperature. TEM images are taken by a JEOL JEM-1011 transmission electron microscope, operated at an acceleration voltage of 100 kV.

### 2.6. Atomic force microscopy (AFM)

The dispersions are diluted 1:30 in Acetone. 100 μL of the dilutions were drop-cast onto Si/SiO$_2$ wafers and dried at 50 °C overnight. AFM images are acquired with a Bruker Innova® AFM in tapping mode using silicon probes (frequency = 300 kHz, spring constant = 40 Nm$^{-1}$) in air. Thickness statistics is performed by measuring ~50 flakes from the AFM images. Statistical analyses are fitted with log-normal distributions. Statistical analyses are performed in WSxM Beta 4.0 software.

### 2.7. Raman spectroscopy

The as-prepared dispersions were drop-cast onto a Si wafer (LDB Technologies Ltd.) covered with 300 nm thermally grown SiO$_2$. Raman measurements are carried out with a Renishaw inVia spectrometer using a 50× objective (numerical aperture 0.75), a laser with a wavelength of 514.5 nm with an incident power of ~1 mW, and a resolution of 1 cm$^{-1}$. A total of 20 points per sample are measured to perform the statistical analyses.

The Raman mappings are carried out on 0.25 cm$^2$ squares, using a 50× objective (numerical aperture 0.75), a laser with a wavelength of 514.5 nm with an incident power of ~5 mW over an area of 20 × 20 μm$^2$. The spot laser diameter was 1 μm; the distance between the measured points was set to 1.5 μm. The mapping was performed quantifying the presence or absence of the $A_g^2$ band of black phosphorus using WiRE 4.4 software from Renishaw.

### 2.8. Fourier Transform infrared spectroscopy (FTIR)

Spectra were collected at ambient conditions on dry samples (powder, ca 1 mg per sample; composites, cut samples of 5 × 5 mm$^2$), using a Fourier Transform Infrared Spectrometer (FTIR; Equinox 70 FT-IR, Bruker) equipped with an attenuated total reflectance accessory (ATR; MIRacle ATR, PIKE Technologies). All spectra are recorded in the range from 4000 to 550 cm$^{-1}$, with a resolution of 4 cm$^{-1}$, accumulating 64 scans.

### 2.9. Optical extinction spectroscopy

Measurements were carried out using a Cary Varian 6000i UVvis-NIR spectrometer using quartz glass cuvette with a path length of 1 cm. Black phosphorus flakes dispersions obtained by WJM were diluted 1:10 with acetone before the measurements. The acetone baseline was subtracted to the as-acquired spectrum.

### 2.10. X-ray photoelectron spectroscopy

Measurements were carried out using a Kratos Axis UltraDLD spectrometer on samples drop-cast in nitrogen onto gold-coated silicon wafers. The XPS spectra were acquired using an Al Kα source operating at 20 mA and 15 kV. The analyses were carried out on a 300 × 700 μm$^2$ area. Wide spectra were acquired at pass energy of 160 and energy step of 1 eV. High-resolution spectra of P 2p, O 1s, C 1s and Au 4f peaks were collected at pass energy of 10 eV and energy step of 0.1 eV. Energy calibration was performed, setting the Au 4f$_{7/2}$ peak at 84.0 eV. Data analysis was carried out with CasaXPS software (version 2.3.17).

### 2.11. Thermogravimetric analysis (TGA).

10–15 mg of composites were cut from the films and analysed by TGA and differential thermal gravimetric analysis under a nitrogen atmosphere (TGA Q500 V20.13 Build 39), using a heating ramp of 10 °C min$^{-1}$ from room temperature to 600 °C. Before the TGA measurement, the samples were dried under vacuum at 60 °C for 12 h.

### 2.12. Tensile tests.

Mechanical properties of composite films were evaluated using a universal testing machine (Instron Dual Column Tabletop 3365), with 5 mm min$^{-1}$ cross-head speed. The tensile measurements were carried out on five different specimens for each film according to ASTM D882 Standard Test Methods for tensile properties of thin plastic sheeting [49].

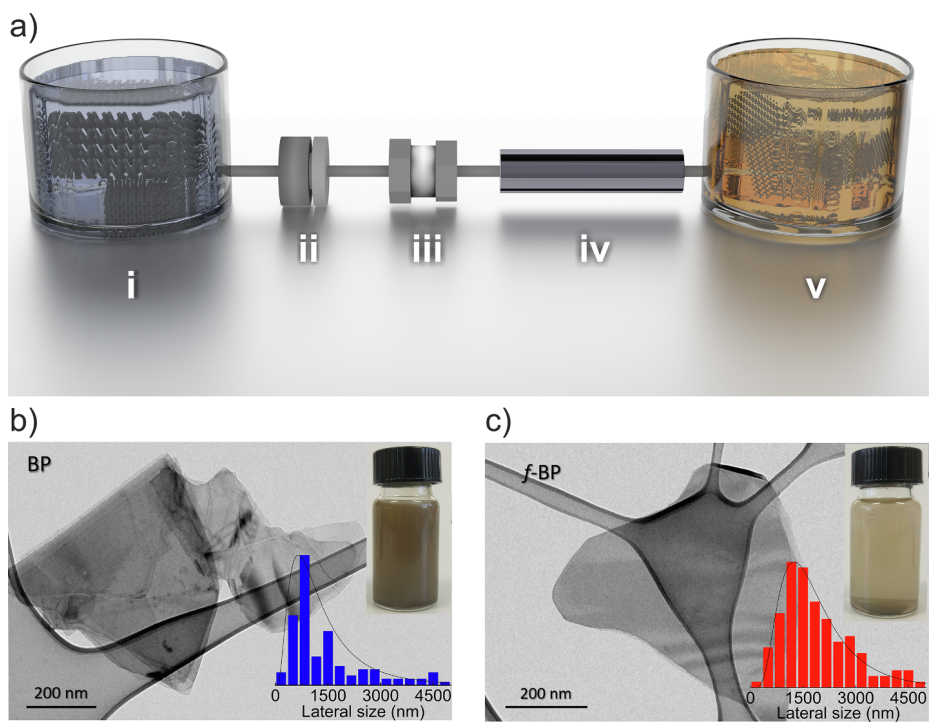

**Fig. 2.** Schematic representation of the wet-jet mill, (i) layered crystal and solvent container, (ii) high-pressure piston, (iii) processor, (iv) chiller and (v) exfoliated 2D crystals dispersion container. (b and c) Image of selected few-layers black phosphorus (BP) and functionalized-few-layers black phosphorus (*f*-BP) flakes, respectively. The histogram as insets show the lateral size distribution where the maximum population (40% and 22%) are found at 642 nm (log. normal s.d. of 0.70) and 1240 nm a (log. normal s.d. of 0.53), for BP and *f*-BP, respectively. Pictures of the dispersions are shown in the up-right insets for the BP and *f*-BP samples.

## 3. Results and discussion

The Wet-jet milling – assisted exfoliation of black phosphorus in acetone produces few-layers black phosphorus dispersions. The detailed exfoliation process of the WJM is reported in the Section 2 and schematised in Fig. 2a. To evaluate the effectiveness of the exfoliation it is necessary firstly to establish a set of figures of merit (FoM), *i.e.*, the exfoliation yield by weight – $Y_W$ [%] –, given by the ratio between the weight of the treated material and the initial weight of the bulk black phosphorus [50]. Additionally, targeting large-scale production of 2D crystals, we previously proposed to set the 1 g of exfoliated crystals as a standard for the definition of two FoM [43]. The first one is the time required to obtain 1 g of exfoliated 2D crystals in dispersion, $t_{1gram}$ [min], and the second one is the volume of solvent required to produce 1 g of exfoliated 2D crystals, $V_{1gram}$ [L]. The quantity $V_{1gram}$ is calculated directly from the concentration of exfoliated flakes in suspension and the $Y_W$. $V_{1gram}$ is a straightforward FoM to evaluate the amount of solvent required for the production of 2D crystals, which is an essential factor in view of large-scale production. For the exfoliation of black phosphorus using the WJM, we found $Y_W$ of ~100%, $t_{1gram}$ of 1.22 min, and $V_{1gram}$ of 0.1 L. These results demonstrate the high production rate of material and that the WJM technique can also be used for the exfoliation of black phosphorus at semi-industrial levels.

Transmission electron microscopy images of representative few-layers black phosphorus (BP) flakes obtained by the WJM are shown in Fig. 2b. Morphological statistical analysis, reported in the inset to Fig. 2b, evidences a flakes' lateral dimension mode of 640 nm (log-normal standard deviation 0.71). The black phosphorus exfoliated in the presence of BBD (*f*-BP, Fig. 2c) shows a lateral size mode of 1240 nm (lognormal standard deviation of 0.53). Similarly, AFM evidences flakes thickness modes of 4.8 (~9 layers) and 3.0 nm (~6 layers) (lognormal standard deviations 1.09 and 1.30), for the BP and *f*-BP respectively, (Fig. S1). Interestingly the size of the *f*-BP is ca. two times the size of the BP, probably the functionalization also mitigates the impact of cavitation and shear rate during the exfoliation process. As a result, the fragmentation process is reduced, obtaining larger flakes, compared to the BP. Additionally, the *f*-BP sample dispersion tends to precipitate, contrary to the BP sample, which is a stable dispersion in acetone, as expected [46]. Pictures of both dispersions are shown as an inset in Fig. 2b and c.

To confirm the structure preservation of the black phosphorus, as well as the protection against oxidation, Raman and infrared spectroscopy measurements are carried out [5,17]. Black phosphorus has 12 phonon modes, *i.e.*, three acoustic and nine optical [51,52]. The Raman active modes are $A_g^1$, $A_g^2$, $B_g$, $B_{2g}$, $B_{3g}^1$ and $B_{3g}^2$ (according to the crystallographic convention where *a* = zigzag, *b* = orthogonal to layers and *c* = armchair, see Fig. 1) [51–57], but only the $A_g^1$, $A_g^2$, and $B_{2g}$ modes are visible in the Raman spectrum when the incident light has no polarization [53], see Fig. 3a. The atomic displacements of the Raman active modes are also shown in Fig. 3a. We performed statistical Raman mapping on random points throughout the samples (the full statistical analysis is detailed in the Supplementary Information Fig. S2). The spectra of bulk black phosphorus (black line), BP (blue line) and *f*-BP (red line) show the characteristic $A_g^1$, $A_g^2$, and $B_{2g}$ Raman peaks. Interestingly, the BP sample shows a decrease in intensity ratio – $I(A_g^1)/I(A_g^2)$ – from 0.84 (s.d. 0.05) for the bulk black phosphorus to 0.48 (s.d. 0.14) for the exfoliated one, Fig. 3b. The decrease of the normalized intensity of $I(A_g^1)/I(A_g^2)$ is ascribed to an increase of the oxygen groups [36,39], *i.e.*, a ratio lower than 0.6 is assigned to oxidized flakes, while, on the contrary, a ratio larger than 0.6 is related to low or no oxidation [36,39]. Contrary to the BP sample, the *f*-BP shows an $I(A_g^1)/I(A_g^2)$ ratio of 0.83 (s.d. 0.1), similar to the bulk counterpart, demonstrating the protection against oxidation due to the functionalization, in agreement with previous works [36,39]. Additionally, the FTIR is performed to confirm the functionalisation and the presence or absence of oxygen groups, see Fig. 3c. The most remarkable difference between BP and *f*-BP spectra is the phosphoric and phosphine oxide families bands (HPO$_4$ and P=O bands at 880 cm$^{-1}$ and 1190 cm$^{-1}$, respectively) [58] present exclusively in the bulk black phosphorus sample. The functional groups Br-aromatic (at 1006 cm$^{-1}$) [58] and –C–P (at 728 cm$^{-1}$) [58] are present exclusively in the *f*-BP sample, confirming the bonding of BBD molecules to the BP flakes. Moreover, the absence of the P=O stretching vibration (1190 cm$^{-1}$) in the *f*-BP, is another indication that further corroborates the protection against oxidation of the *f*-BP sample. X-ray photoelectron spectroscopy is carried out to confirm the P–C and Br–C bonding, and the presence or absence of oxygen groups.

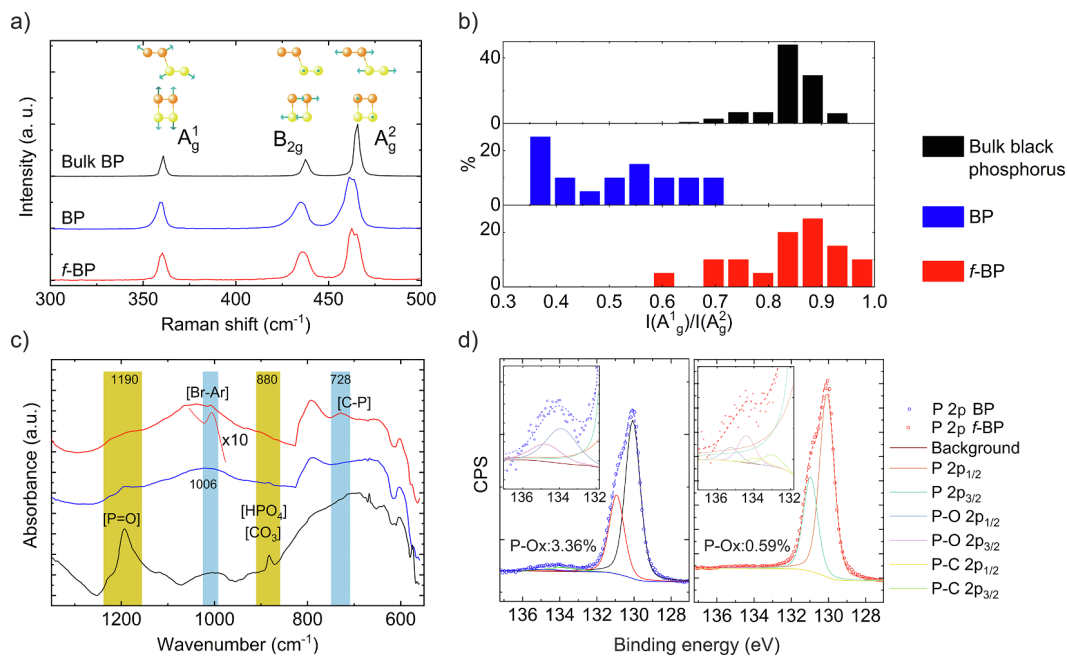

**Fig. 3.** Spectroscopy characterisation of the bulk black phosphorus in black, BP in blue and *f*-BP in red lines/bars. (a) Raman spectra of the three samples, the inset shows the Raman-active vibrational modes for their corresponding peaks. (b) Statistical analyses of the $A_g^1/A_g^2$ ratios. (c) Normalised Fourier-Transform infrared spectra of the three samples, the amber rectangle symbolises the functional groups present just in the bulk black phosphorus, and in blue, the groups present only in the *f*-BP. The peak at 1006 cm$^{-1}$ in the *f*-BP sample is ten-fold magnified. (d) X-ray photoelectron spectroscopy of the O1s and P 2p peaks obtained from the two samples.

Fig. 3d shows the $P_{2p}$ region for the BP and *f*-BP samples, with the main contribution at ≈130 eV coming from phosphorus in the black phosphorus structure. It is noteworthy the presence of a band between 136 and 134 eV in the BP sample. The inset shows the peaks related to oxidised phosphorus (P–O groups) present in the 3.4%. Contrary, in the *f*-BP (right plot) the peaks are visible zooming only after the 136–134 eV zone, and the deconvolution reveals the presence of P–O groups in a 0.59%. Furthermore, the presence of two extra bands representing the P–C bonding is observed. The presence of Br is exclusive for the *f*-BP sample, see the supplementary information (S3)

It has been demonstrated that large-sized flakes provide the best mechanical improvement in polymer composites [59]. More relevant, the as-produced few-layers graphene flakes were applied as fillers in polyamide matrix [43], enhancing by 34% the flexural modulus in the composite graphene-polyamide, demonstrating that it is possible to use un-purified flakes processed by the wet-jet mill, where the large flakes may play an essential role in the improvement of the mechanical performance. Consequently, the as-produced BP and *f*-BP dispersions (*i.e.*, the dispersions without flake size sorting through centrifugation) are mixed with polymethyl methacrylate (PMMA) for the realisation of the composites, see the Section 2 for detailed information.

The PMMA is a transparent thermoplastic polymer; it has a tensile strength of 76 ± 0.5 MPa [34], an impact resistance of 20.7 ± 1.2 kJm$^{-2}$ [34], a low optical loss in the visible spectrum and is thermally stable (melting point 160 °C) [34]. These properties make PMMA an interesting substrate for opto- and tribo-electronics and for the design/realisation of flexible and wearable devices [29,43]. Fig. 4a shows the images of the composites with BP and *f*-BP/PMMA mass ratios, *i.e.*, 0.01, 0.05, 0.1, 0.2 and 0.5 wt%, obtaining the samples BP0.01, BP0.05, BP0.1, BP0.2, BP0.5 and *f*-BP0.01, *f*-BP0.05, *f*-BP0.1, *f*-BP0.2, *f*-BP0.5, respectively.

The optical transmittance, see Fig. 4b, decreases with the increase of the concentration of BP and *f*-BP. Specifically, at 600 nm, the transmittance of the PMMA film is 87%, decreasing to 70% with BP0.01 and *f*-BP0.01, and further decrease to 21% and 9% with BP0.5 with *f*-BP0.5, respectively. The difference in the optical transmittance is attributed to the better distribution of the BP flakes into the polymeric matrix. The stability of the exfoliated flakes in the solvent may also play an important role in the polymer matrix dispersions. The presence of functional groups on the surface of the flake changes the chemistry of the samples, promoting the aggregation in the solvent.

The level of oxidation of the as-prepared composites is characterised by FTIR (Fig. 4c). Weak FTIR peaks related to P–O and/or P=O families [58] at 620 cm$^{-1}$ are present just in the BP/PMMA sample (enlarged 10 times for a clear visualisation) demonstrating that the functionalization is an effective method to protect the BP flakes against oxidation, as previously reported [8,34,35]. The thermogravimetric analysis, Fig. S4, also shows an increase in thermal stability. The thermal degradation of the PMMA shows a characteristic one-step decomposition with an onset temperature at 357 °C [60], while the composite degradation occurs at higher temperatures when the fillers are present, *i.e.*, at 367 °C, for both BP and *f*-BP samples. Additionally, the presence of the functional groups is also observed comparing the weight loss of BP and *f*-BP powders with the bulk black phosphorus, showing a weight-loss difference of 2.7% and 4.8% at 350 °C. The difference in weight loss indicates the presence of 2.1 wt% of functional groups in the *f*-BP sample.

We studied the distribution of BP flakes into the PMMA matrix by micro Raman mappings (Fig. 4d) and energy-dispersive X-ray – scanning electron microscopy (EDS-SEM, Fig. 4e). Raman mappings are carried out on areas of 30 × 30 μm$^2$, focusing on the presence or the absence of the $A_g^2$ band. The mapping image is the result of the normalisation of the intensity of the $A_g^2$ band – $I(A_g^2)$ – with the υ(CH$_2$) stretching mode of PMMA band at 812 cm$^{-1}$ – I(PMMA$_{@812}$) –. In the Raman mapping image, the dark blue (for the BP/PMMA composites) or dark red (for the *f*-BP/PMMA composites) tonalities represent the presence of flakes in the composite. Contrary, a light-blue (for the BP/PMMA composites) or light-red (for the *f*-BP/PMMA composites) tonalities represent the absence of flakes in the composites. The two Raman mappings demonstrate the presence of flakes into the polymer matrix, also showing that the aggregation of flakes rises with their concentration. Despite this fact, it is noteworthy that the Raman mapping analysis confirms that the pressing process does not damage the crystal structure of the flakes, assuring its integrity after the polymer blending [5,29,61]. Additionally, the SEM characterisation gives us a direct visualisation of the flakes embedded into the PMMA matrix. Figs. 4e and S5 confirm the presence of BP (i) and *f*-BP flakes (ii) embedded into the polymer matrix, *i.e.*, the BP0.5 sample presents BP aggregates in the polymer matrix. However, the aggregation of the fillers at loadings of 0.5 wt% is a recurrent phenomenon in polymer/2D crystals composites [62]. The elemental mapping shows the homogenous presence of oxygen species in the scanned area (Fig. 4e–i), coming from both the PMMA and the BP flakes. Contrary, for the *f*-BP0.5 it is notable, the absence of oxygen species on the flakes surfaces, also demonstrating the oxygen protection of the BP flakes upon functionalization.

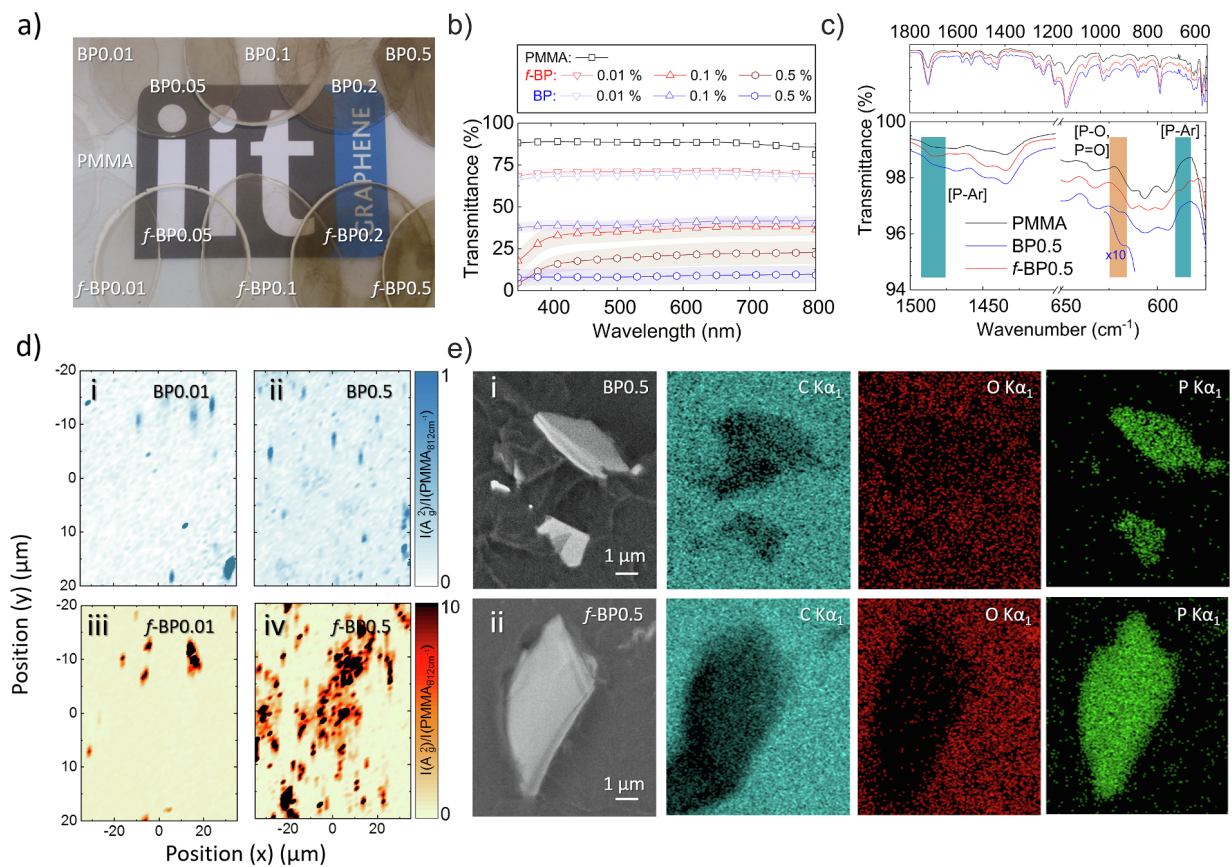

**Fig. 4.** (a) Picture of the bare PMMA, BP/PMMA and *f*-BP/PMMA composites with different filler loadings. Spectroscopy of the as-prepared films with 0.01, 0.1 and 0.5 wt% of filler by (b) optical transmittance and (c) FTIR. (d) Raman mapping on the composite films, the figure shows the mappings in blue shades for the composites at 0.01 (i), and 0.5 wt% (ii) BP mass loadings and in red shades the 0.01 (iii) and 0.5 wt% (iv) *f*-BP mass loadings. The presence of black phosphorus flakes is represented by the normalised intensity of the $A_g^2$ peak with the PMMA peak at 812 cm$^{-1}$. (e) Energy-dispersive X-ray – scanning electron microscopy images of cross-sections of the composites BP0.5 (i) and *f*-BP0.5 (ii), the carbon, oxygen, and phosphorus elemental mappings are shown in cyan, red and green respectively.

The possible degradation of the exfoliated black phosphorus due to ambient exposure was analysed by Raman spectroscopy since the Raman signal of black phosphorus (crystalline integrity) is sensitive to the presence of oxygenated species [31]. Fig. S6 shows the Raman mapping of the BP/ and *f*-BP/PMMA composites after eight weeks of ageing under ambient conditions, demonstrating the stability of the flakes against oxidation into the polymer matrix. However, the BP/ PMMA composite show broad Raman peaks (FWHM from 10 to 20 cm$^{-1}$), indicating an on-going degradation [31]. Contrariwise, the *f*-BP/PMMA composite shows narrow Raman peaks (FWHM from 4 to 6 cm$^{-1}$).

The enrichment of polymers with semiconducting nanomaterials could improve its physical [63] and chemical properties [63,64], boosting the commercialisation of polymer composites for several applications, *e.g.*, in capacitors [65], batteries [46,66], and sensors [16,33]. Worth noting, the addition of nanomaterials into a polymer matrix could modify the mechanical performances of the polymer itself, which is an important property to consider especially for applications in which the mechanical stress is persistent, *e.g.*, in flexible, stretchable and wearable devices, as well as in actuators, and tribology. Therefore, the mechanical properties of the as-prepared *f*-BP/PMMA and BP/ PMMA composites are analysed via tensile testing. This test consists in applying a constant uniaxial tensile stress to the samples, (cut with a known geometrical shape), the force required for the elongation and fracture in the composite is expressed as load (or force) plotted against the deformation. The parameters that are obtained by the tensile testing are the slope of the curve in the elastic region (Young modulus, E), the ultimate tensile strength and the elongation at break [67,68].

The stress vs strain curves of pristine PMMA (grey curve) and composites at different BP and *f*-BP loadings are shown in Fig. 5 and summarised in Table 1. The E reaches a maximum value of 2.92 ± 0.28 GPa with BP0.01, and the modulus decreases with the increase of the BP filler, reaching the 2.40 GPa with BP0.5 corresponding to 69% improvement in comparison to pristine PMMA. However, in the case of *f*-BP, the highest E = 2.71 ± 0.17 GPa is observed with *f*-BP0.05, representing an increment of 90.8% compared to pristine PMMA, (E = 1.42 ± 0.01 GPa). Increasing the *f*-BP fillers content the E values drop, reaching a minimum value of 2.06 GPa at 0.5 wt% corresponding to 45% improvement compared to the value of the pristine PMMA.

The decreases of E with the loading of few-layers black phosphorus superior in *f*-BP0.01 and *f*-BP0.05% is associated to flakes aggregation [69–71], as corroborated by SEM and Raman mappings shown in Fig. 4.

The presence of few-layers black phosphorus improves the ultimate tensile strength ($\sigma_U$), defined as the higher stress value reached in the stress vs strain curve. Similarly, to the results obtained for E, the highest $\sigma_U$ is ~71 MPa with BP0.01 and ~64 MPa with *f*-BP0.05. Moreover, there is a sudden decrease in $\sigma_U$ observed, which is attributed to the aggregation of the fillers.

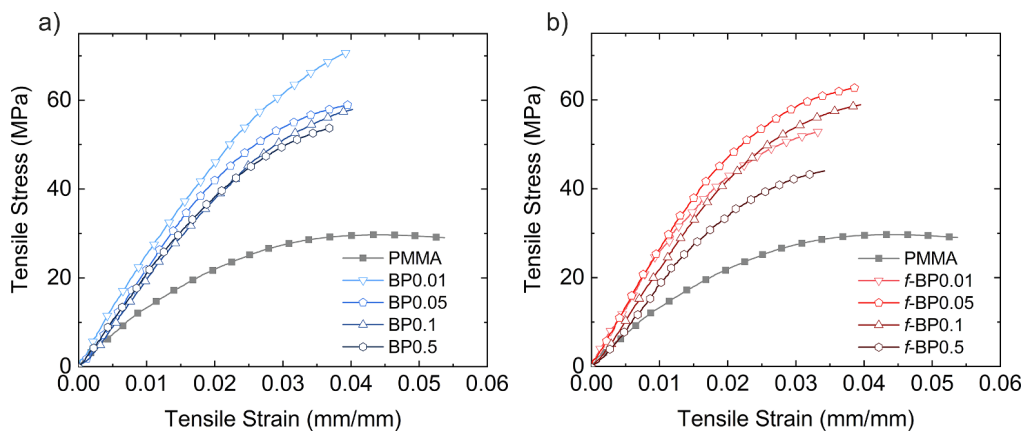

**Fig. 5.** Mechanical properties of composites and pristine materials: (a) using the BP samples and (b) the *f*-BP samples as reinforcement materials.

**Table 1**
Mechanical performance of the as-prepared few-layers black phosphorus-based composites.

| Sample | Black phosphorus %wt | Young's Modulus (E) (GPa) | (std. Dev.) | Ultimate Tensile Strength ($\sigma_U$) (MPa) | (std. Dev.) |
|---|---|---|---|---|---|
| PMMA | 0 | 1.42 | 0.01 | 30.38 | 1.44 |
| BP | 0.01 | 2.92 | 0.28 | 71.15 | 8.18 |
|  | 0.05 | 2.76 | 0.02 | 59.42 | 4.46 |
|  | 0.1 | 2.66 | 0.14 | 59.48 | 2.63 |
|  | 0.5 | 2.40 | 0.28 | 53.11 | 4.37 |
| *f*-BP | 0.01 | 2.46 | 0.38 | 52.89 | 4.51 |
|  | 0.05 | 2.71 | 0.17 | 63.78 | 2.90 |
|  | 0.1 | 2.58 | 0.20 | 59.96 | 2.74 |
|  | 0.5 | 2.06 | 0.02 | 44.40 | 1.12 |

The mechanical test demonstrates that the inclusion of exfoliated flakes of black phosphorus improves the overall mechanical performances of PMMA. Despite this fact, the BP/PMMA composite outperformed mechanically the *f*-BP/PMMA one. From the chemical point of view, the best stability against environmental oxidation is obtained by the *f*-BP/PMMA sample. An optimal compromise shall be found balancing the chemical with mechanical performances to obtain the best mechanical and chemical improvements. Furthermore, this work comprises the black phosphorus into the family of 2D crystals, which enhances the mechanical properties of polymers [10,19,20,29,72,73].

## 4. Conclusions

We demonstrated the exfoliation of black phosphorus using a semi-industrial scale approach, *i.e.*, the wet-jet milling. This technique enables the production of few-layers black phosphorus (BP) dispersions with a concentration of 0.5 g L$^{-1}$. The exfoliated flakes present a lateral size and thickness modes of 640 nm and 4.8 nm, respectively. Furthermore, we demonstrated the single step exfoliation and functionalization of black phosphorus (*f*-BP) with the wet-jet mill. To this end, we used 4-bromobenzene diazonium tetrafluoroborate to functionalize the black phosphorus. The *f*-BP presents a lateral size and thickness modes of 1200 nm and 3.0 nm respectively, and a functional group every 250 phosphorus atoms. The functionalisation facilitated the dispersion into the polymer matrix and also prevented the degradation of the BP flakes due to oxidation.

Additionally, we designed and produced a composite of polymethyl methacrylate with BP and *f*-BP. The composites with BP and *f*-BP increased the polymethyl methacrylate Young's moduli us by ~106% and ~91%, respectively. Our results demonstrate the reliability of exfoliated and functionalised black phosphorus as reinforcement filler for composites, pushing its exploitation in applications in which the embedding into polymer matrixes, *e.g.*, in triboelectric or wearable devices, represents a real benefit.

## Acknowledgements

This project has received funding from the European Union's Horizon 2020 research and innovation programme under the grant agreement No. 785219 – GrapheneCore2. The authors thank Paraskevi Papadopoulou for the FTIR measurements and Giammarino Pugliese for the TGA measurement, and Simone Lauciello for the EDS-SEM images.

## Appendix A. Supplementary data

Supplementary data to this article can be found online at https://doi.org/10.1016/j.flatc.2019.100131.